\documentclass[11pt,a4paper,showpacs,showkeys,,superscriptaddress]{article}

\usepackage{epsfig}
\usepackage{amsmath,amssymb}
\usepackage{graphicx}
\usepackage{xcolor}
\usepackage{subfigure}

\usepackage{latexsym}
\usepackage{hyperref}
\hypersetup{colorlinks,%
  citecolor=blue,%
  linkcolor=red,%
  pdftex}

\begin{document}

\baselineskip 6mm
\renewcommand{\thefootnote}{\fnsymbol{footnote}}

\newcommand{\nc}{\newcommand}
\newcommand{\rnc}{\renewcommand}



\newcommand{\tcb}{\textcolor{blue}}
\newcommand{\tcr}{\textcolor{red}}
\newcommand{\tcg}{\textcolor{green}}


\def\beq{\begin{equation}}
\def\eeq{\end{equation}}
\def\ba{\begin{array}}
\def\ea{\end{array}}
\def\bea{\begin{eqnarray}}
\def\eea{\end{eqnarray}}
\def\nn{\nonumber}


\def\CMP{Commun. Math. Phys.~}
\def\JHEP{JHEP~}
\def\Pre{Preprint}
\def\PRL{Phys. Rev. Lett.~}
\def\PR {Phys. Rev.~}
\def\CQG {Class. Quant. Grav.~}
\def\PL {Phys. Lett.~}
\def\NP {Nucl. Phys.~}

\def\G{\Gamma}

\def\S{{\bf S}}
\def\C{{\bf C}}
\def\Z{{\bf Z}}
\def\R{{\bf R}}
\def\N{{\bf N}}
\def\M{{\bf M}}
\def\P{{\bf P}}
\def\bm{{\bf m}}
\def\bn{{\bf n}}

\def\CA{{\cal A}}
\def\CB{{\cal B}}
\def\CC{{\cal C}}
\def\CD{{\cal D}}
\def\CE{{\cal E}}
\def\CF{{\cal F}}
\def\CH{{\cal H}}
\def\CM{{\cal M}}
\def\CG{{\cal G}}
\def\CI{{\cal I}}
\def\CJ{{\cal J}}
\def\CL{{\cal L}}
\def\CK{{\cal K}}
\def\CN{{\cal N}}
\def\CO{{\cal O}}
\def\CP{{\cal P}}
\def\CQ{{\cal Q}}
\def\CR{{\cal R}}
\def\CS{{\cal S}}
\def\CT{{\cal T}}
\def\CU{{\cal U}}
\def\CV{{\cal V}}
\def\CW{{\cal W}}
\def\CX{{\cal X}}
\def\CY{{\cal Y}}
\def\CZ{{\cal Z}}

\def\We{{W_{\mbox{eff}}}}


\newcommand{\Lie}{\pounds}

\newcommand{\p}{\partial}
\newcommand{\bp}{\bar{\partial}}

\newcommand{\half}{\frac{1}{2}}

\newcommand{\bfalpha}{{\mbox{\boldmath $\alpha$}}}
\newcommand{\bfbeta}{{\mbox{\boldmath $\beta$}}}
\newcommand{\bfgamma}{{\mbox{\boldmath $\gamma$}}}
\newcommand{\bfmu}{{\mbox{\boldmath $\mu$}}}
\newcommand{\bfpi}{{\mbox{\boldmath $\pi$}}}
\newcommand{\bfvarpi}{{\mbox{\boldmath $\varpi$}}}
\newcommand{\bftau}{{\mbox{\boldmath $\tau$}}}
\newcommand{\bfeta}{{\mbox{\boldmath $\eta$}}}
\newcommand{\bfxi}{{\mbox{\boldmath $\xi$}}}
\newcommand{\bfkappa}{{\mbox{\boldmath $\kappa$}}}
\newcommand{\bfepsilon}{{\mbox{\boldmath $\epsilon$}}}
\newcommand{\bfTheta}{{\mbox{\boldmath $\Theta$}}}

\newcommand{\bz}{{\bar{z}}}

\newcommand{\dalpha}{\dot{\alpha}}
\newcommand{\dbeta}{\dot{\beta}}
\newcommand{\blambda}{\bar{\lambda}}
\newcommand{\btheta}{{\bar{\theta}}}
\newcommand{\bsigma}{{{\bar{\sigma}}}}
\newcommand{\bepsilon}{{\bar{\epsilon}}}
\newcommand{\bpsi}{{\bar{\psi}}}


\def\ct{\cite}
\def\la{\label}
\def\eq#1{(\ref{#1})}


\def\a{\alpha}
\def\b{\beta}
\def\g{\gamma}
\def\G{\Gamma}
\def\d{\delta}
\def\D{\Delta}
\def\ep{\epsilon}
\def\e{\eta}
\def\ph{\phi}
\def\Ph{\Phi}
\def\ps{\psi}
\def\Ps{\Psi}
\def\k{\kappa}
\def\l{\lambda}
\def\L{\Lambda}
\def\m{\mu}
\def\n{\nu}
\def\th{\theta}
\def\Th{\Theta}
\def\r{\rho}
\def\s{\sigma}
\def\S{\Sigma}
\def\ta{\tau}
\def\o{\omega}
\def\O{\Omega}
\def\pr{\prime}


\def\half{\frac{1}{2}}

\def\goto{\rightarrow}

\def\na{\nabla}
\def\grad{\nabla}
\def\curl{\nabla\times}
\def\div{\nabla\cdot}
\def\pa{\partial}

\def\bra{\left\langle}
\def\ket{\right\rangle}
\def\lb{\left[}
\def\lc{\left\{}
\def\ls{\left(}
\def\lp{\left.}
\def\rp{\right.}
\def\rb{\right]}
\def\rc{\right\}}
\def\rs{\right)}
\def\cl{\mathcal{l}}

\def\vac#1{\mid #1 \rangle}

\def\td#1{\tilde{#1}}
\def\check{ \maltese {\bf Check!}}


\def\Tr{{\rm Tr}\,}
\def\det{{\rm det}\,}


\def\bc#1{\nnindent {\bf $\bullet$ #1} \\ }
\def\ch {$<Check!>$ }
\def\ss {\vspace{1.5cm}}

\begin{titlepage}

\hfill\parbox{5cm} { }

\hskip1cm

\vspace{10mm}

\begin{center}
{\Large \bf Frame-independent holographic conserved charges}

\vskip 1. cm
  { Seungjoon Hyun\footnote{e-mail : sjhyun@yonsei.ac.kr}, Jaehoon Jeong\footnote{e-mail : jjeong@physics.auth.gr}, Sang-A Park\footnote{e-mail : sangapark@yonsei.ac.kr},
  Sang-Heon Yi\footnote{e-mail : shyi@yonsei.ac.kr} 
  }

\vskip 0.5cm

{\it Department of Physics, College of Science, Yonsei University, Seoul 120-749, Korea}\\
{\it ${}^{\dagger}$Institute of Theoretical Physics,
Aristotle University of Thessaloniki,  Thessaloniki
54124, Greece}
\end{center}

\thispagestyle{empty}

\vskip1.5cm

 
\centerline{\bf ABSTRACT} \vskip 4mm
We propose the modified form of the conventional holographic conserved charges which provides us the frame-independent expressions for charges. This form is also shown to be independent of  the holographic renormalization scheme.  We show the frame and scheme independence through the matching of our holographic expression to the covariant bulk expression of conserved charges. As an explicit example, we consider five-dimensional AdS Kerr black holes and show that our form of holographic conserved charges gives us the identical expressions in the rotating and non-rotating frames. \vspace{1cm} 
\noindent 
\vspace{2cm}


\end{titlepage}

\renewcommand{\thefootnote}{\arabic{footnote}}
\setcounter{footnote}{0}

\section{Introduction}

Holographic principle in modern physics has been introduced as the fundamental property of quantum gravity, which was speculated on the basis of the area nature of the black hole entropy. After its concrete realization in the form of the AdS/CFT correspondence, it becomes one of main research arena and has been studied in various contexts. Especially, the AdS/CFT correspondence has been used as a modern toolkit of strong coupling phenomena for the dual field theory. In this context holography has many interesting applications and implications even at the level of a classical theory of gravity, since the classical computation in gravity has the dual interpretation for quantum phenomena in the field theory side. Conversely, it also provides new approaches to the classical theory of gravity through the perspective from the dual field theory. One such application is the introduction of holographic approach to conserved charges in the classical theory of gravity which have been explored in the huge number of literatures. 

Holographic conserved charges in the asymptotic AdS space~\cite{Balasubramanian:1999re} are introduced along with the construction of boundary stress tensor in gravity by using the Brown-York formalism~\cite{Brown:1992br}, which is now regarded as one of the AdS/CFT dictionary. Despite their successful applications to various cases, holographic charges need to be compared and/or matched to traditional bulk charges since their equivalence is not warranted a priori. In Einstein gravity with negative cosmological constant, the equivalence between the holographic and traditional bulk  conserved  charges of black holes are shown in Ref.s~\cite{Hollands:2005wt,Papadimitriou:2005ii,Hollands:2005ya}.  Interestingly, it was observed that holographic conserved charges of black holes might be different from those by the covariant phase space method when the conformal anomaly of the dual field theory does not vanish.  In particular, it has been noticed that the results from the conventional expression of holographic charges depends on the frames at the asymptotic AdS space in odd dimensions, while the charges in covariant phase space method remain invariant. When the metric for the asymptotic AdS space in odd dimensions is taken in the standard non-rotating  form, the Casimir energy is given just by constant. On the other hand, the Casimir energy becomes dependent on the rotational parameters when the metric is taken in the rotating frame~\cite{Awad:1999xx,Papadimitriou:2005ii,Gibbons:2005jd}. Furthermore,  the conventional expression for holographic charges depends on the counter term subtraction scheme~\cite{deHaro:2000xn,Bianchi:2001kw}.

Since it was shown that conserved  charges by the covariant phase space method should be completely consistent with the first law of black hole thermodynamics~\cite{Wald:1993nt},  the difference between holographic and covariant phase space charges means that conserved charges by the holographic method require the modification of the first law of black hole thermodynamics, albeit the minimal modification of the first law is shown to be sufficient for harmless physical interpretation of holographic results~\cite{Papadimitriou:2005ii}. Still it would be nice if there is a construction of holographic charges in such a way that they are identical  with the bulk ones and thus satisfy the standard form of the first law of black hole thermodynamics.

In this paper we would like to revisit the construction of the conventional holographic conserved charges and show how it can be modified to give identical results with the bulk constructions. Our approach is based on the recent works~\cite{Kim:2013zha,Kim:2013cor,Hyun:2014kfa,Gim:2014nba,Hyun:2014sha} which can be regarded as the generalization of the traditional Abbott-Deser-Tekin(ADT) formalism~\cite{Abbott:1981ff,Abbott:1982jh,Deser:2002rt,Deser:2002jk} to the holographic setup.  It turns out that our construction is rather general and completely consistent with the bulk covariant expression of conserved charges under a very mild assumption. As a result, whenever the boundary stress tensor is well-defined and there is a continuous parameter in the black hole solution,  our expression of holographic charges gives finite, frame and scheme independent results and is completely consistent with the standard form of the first law of black hole thermodynamics. 


\section{Modified holographic conserved charges}

Let us start from the brief summary of holographic renormalization in this section. See~\cite{Skenderis:2002wp} for a review.
In terms of the boundary values $(\gamma, \psi)$ of the bulk metric and matter fields $\Psi\equiv (g,\psi)$, the on-shell renormalized  action is given by (See the Ref.~\cite{Hyun:2014sha} for our convention)
\[   
I^{on}_{r}[\gamma, \psi] = I [g, \psi]_{\rm on-shell} + I_{GH}[\gamma] + I_{ct}[\gamma,\psi]\,,
\]
where the Gibbons-Hawking and counter terms  $I_{GB}, I_{ct}$ are defined on a hypersurface. The on-shell condition renders the renormalized action $I_{r}^{on}$ to be the functional of the boundary value $(\gamma, \psi)$ at the boundary $\CB$. The generic variation of the on-shell renormalized action is taken in the form of 
\begin{equation} \label{ReAction}
\delta I^{on}_{r}[\gamma, \psi] = \frac{1}{16\pi G}\int_{\CB} d^{d} x~ \sqrt{-\gamma}\Big[ T_{B}^{ij}\delta \gamma_{\,ij} + \Pi_{\psi}\delta \psi\Big]\,.
\end{equation}

In order to introduce the boundary ADT current in the renormalized boundary action, let us recall that the boundary diffeomorphism results in the  identity  of the form:
\begin{equation} \label{Bid}
\nabla_{i}({2\bf T}^{ij}_{B}\zeta^{B}_{j})=T^{ij}_{B}\, \Lie_{\zeta_{B}}\gamma_{ij} + \Pi_{\psi}\, \Lie_{\zeta_{B}}\psi\,,
\end{equation}
where $\Lie_{\zeta_{B}}$ denotes the Lie derivative on the boundary and ${\bf T}^{ij}_{B}$ does the modified boundary stress tensor defined by 
\[   
{\bf T}^{ij}_{B} \equiv T^{ij}_{B} + \frac{1}{2}{\cal Z}^{ij}_{B}\,, \qquad T^{ij}_{B} \equiv \frac{1}{\sqrt{-\gamma}}\frac{\delta I^{on}_{r}}{\delta \gamma_{ij}}\,.
\]
The above boundary identity can be regarded as the analog of the bulk Noether identity, of which elementary derivation is given in~\cite{Hyun:2014sha}. Note that ${\cal Z}$-tensor does not need to be a symmetric one and is given in terms of $\Pi_{\psi}$'s.

Let us introduce the boundary conserved current as
\begin{align}   \label{Bcur}
 \CJ^{i}_{B} (\xi_B) \equiv&  - \,\delta{\bf T}^{ij}_{B}  \xi^{B}_{j} - \frac{1}{2}\gamma^{kl}\delta\gamma_{kl} {\bf T}^{ij}_{B}  \xi^{B}_{j} - {\bf T}^{ij}_{B}  \delta\gamma_{jk}\xi_{B}^{k} + \frac{1}{2}\, \xi^{i}_{B}\big(T^{kl}_{B}\delta \gamma_{kl} + \Pi_{\psi}\delta \psi\big)\,,
\end{align}
where $\delta$ denotes a linearization with respect to the boundary fields, including the variations of Killing vectors.  This current can be written in the form of
\begin{align}   \label{}
\label{Bcur1}
\sqrt{-\gamma} \CJ^{i}_{B} (\xi_B) =& - \delta \Big(\sqrt{-\gamma}\,{\bf T}^{ij}_{B}\xi^{B}_{j}\Big) + \sqrt{-\gamma}\, {\bf T}^{i}_{\!B\, j}\delta \xi^{j}_{B} + \frac{1}{2}\, \sqrt{-\gamma}\xi^{i}_{B}\big(T^{kl}_{B}\delta \gamma_{kl} + \Pi_{\psi}\delta \psi\big)\,.
\end{align}
One may note that the first term corresponds to the linearized form of the conserved currents in conventional holographic charges.  For a boundary Killing vector $\xi_{B}$,  the conservation of the first term is the simple result of the identity given in Eq.~(\ref{Bid}). Interestingly,  this identity also leads to the conservation of the sum of the second and third terms as shown in the Appendix. After taking the  linearization of the boundary fields along the black hole parameters  and  integrating the linearized form along the one-parameter path $ds$, the holographic charges are introduced  by
\begin{equation} \label{Bcharge}
Q_{B}(\xi_{B}) \equiv \frac{1}{8 \pi G} \int ds \int d^{d-1}x_{i}\sqrt{-\gamma}{\cal J}^{i}_{B} 
\,. 
\end{equation}

We would like to emphasize that our choice of the conserved boundary currents is 
motivated by the bulk off-shell extension of the conventional ADT formalism and its form in~Eq.~(\ref{Bcur}) is already written down in Ref.~\cite{Hyun:2014sha}. Our boundary current in Eq.~(\ref{Bcur1}) is a generalization in the case of boundary Killing vectors varying under a generic variation.  It turns out that this generalization of conserved currents leads to the frame-independent expression of conserved charges, which is also free from the ambiguity in the counter term subtraction. This advantage becomes manifest by showing the equivalence of the boundary currents to the bulk ADT potential expressions for charges, which is given in the following section.

\section{Scheme and frame independence}
In this section we argue that our boundary construction of currents leads to the scheme independent results by showing their equivalence with covariant bulk expression for the ADT potential of conserved charges. To this purpose, we explain how to construct the off-shell ADT potential even when a bulk Killing vector is varied under a generic variation.

In the bulk, there is an off-shell identity known as the Noether identity which can be written in the form of 
\begin{align}   \label{}
\CE_{\Psi}\Lie_{\zeta}\Psi &\equiv \CE_{\mu\nu}\, \Lie_{\zeta} g^{\mu\nu} + \CE_{\psi}\,\Lie_{\zeta}\psi= -2\nabla_{\mu}({\bf E}^{\mu\nu}\zeta_{\nu})\,,  \qquad {\bf E}^{\mu\nu}\equiv \CE^{\mu\nu} + \frac{1}{2}{\cal Z}^{\mu\nu}   \,,
\end{align}
where ${\cal E}_{\Psi} $ denotes the Euler-Lagrange expression for the field $\Psi$ and ${\cal Z}^{\mu\nu}$ tensor is given in terms of matter Euler-Lagrange expressions, $\CE_{\psi}$. 
For a Killing vector $\xi$  which may be unpreserved under a generic variation, one can introduce the off-shell ADT current, just like in the non-varying case~\cite{Hyun:2014sha}  as
\begin{align}  
\label{ADT}
\CJ^{\mu}_{ADT} (\xi, \delta \Psi) =&~ \delta {\bf E}^{\mu\nu}\xi_{\nu} + \frac{1}{2}g^{\alpha\beta}\delta g_{\alpha\beta}\, {\bf E}^{\mu\nu}\xi_{\nu}  + {\bf E}^{\mu\nu}\delta g_{\nu\rho}\, \xi^{\rho} + \frac{1}{2}\xi^{\mu}\CE_{\Psi}\delta \Psi\,,
\end{align}
which can be rewritten as
\begin{align}   \label{}
\label{ADTcpt}
\sqrt{-g}\CJ^{\mu}_{ADT} (\xi, \delta \Psi) =&~ \delta\Big(\sqrt{-g}\, {\bf E}^{\mu\nu}\xi_{\nu}\Big) - \sqrt{-g}\, {\bf E}^{\mu}_{~\nu}\, \delta\xi^{\nu} + \frac{1}{2}\sqrt{-g}\,\xi^{\mu}\CE_{\Psi}\delta \Psi\,.
\end{align}
This expression may be regarded a slight generalization of the non-varying Killing vector case~\cite{Kim:2013zha,Hyun:2014sha}. Note that this  current takes the same structure as the boundary conserved current in the previous section. The off-shell conservation of this current ${\cal J}^{\mu}_{ADT}$ allows us to write this current in terms of the potential as ${\cal J}^{\mu}_{ADT}=\nabla_{\nu}Q^{\mu\nu}_{ADT}$ at the off-shell level. 

For the bulk Killing vector $\xi$,  one can see  that the symplectic current~\cite{Lee:1990nz,Wald:1993nt,Iyer:1994ys} defined for a generic diffeomorphism parameter $\zeta$ by
$
\omega(\Lie_{\zeta}\Psi, \delta \Psi) \equiv \Lie_{\zeta}\Theta^{\mu}(\delta \Psi\,;\, \Psi) - \delta \Theta^{\mu}(\Lie_{\zeta}\Psi\,;\, \Psi)
$,
reduces to 
\begin{equation} \label{}
\omega(\Lie_{\xi}\Psi, \delta \Psi)  = -\Theta^{\mu}(\Lie_{\delta \xi}\Psi\,;\, \Psi)\,,
\end{equation}
wherer $\Theta^{\mu}(\delta \Psi)$ is the surface term for a generic variation of the bulk Lagrangian $\CL$ given by $\delta (\sqrt{-g}\CL) = \sqrt{-g}\CE_{\Psi}\delta \Psi + \p_{\mu}\Theta^{\mu}(\delta \Psi)$.
Through relations among the ADT current, symplectic current and the off-shell Noether current for a diffeomorphism variation  ${\cal J}^{\mu}_{\zeta} \equiv 2\sqrt{-g}{\bf E}^{\mu\nu}\zeta_{\nu} + \zeta^{\mu}\sqrt{-g}\CL - \Theta^{\mu}$, 
the final off-shell expression of the ADT potential, up to the irrelevant total derivative term, turns out to be 
\begin{align}  \label{bulkADT}
2\sqrt{-g} Q^{\mu\nu}_{ADT}(\xi, \delta \Psi\,;\, \Psi) =&~ \delta K^{\mu\nu}(\xi\,;\, \Psi) - K^{\mu\nu}(\delta \xi\,;\,\Psi) -2\xi^{[\mu}\Theta^{\nu]}(\delta \Psi\,;\,\Psi)\,. 
\end{align}
This final expression can be regarded as a slight generalization of covariant phase space results~\cite{Wald:1993nt,Iyer:1994ys}, which has already been obtained in Einstein gravity in~\cite{Barnich:2004uw}. 

The matching between the boundary current ${\cal J}^{i}_{B}$ and the bulk ADT potential $Q^{\mu\nu}_{ADT}$ goes in the same way just as in the case of $\delta \xi^{\mu}=0$ and $\delta \xi^{i}_{B} =0$, as follows. Let us take the Fefferman-Graham coordinates for the asymptotic AdS space as $ds^{2} = d\eta^{2} + \gamma_{ij}dx^{i}dx^{j}$.
Adding the Gibbons-Hawking and counter terms in holographic renormalization gives us the additional surface terms  modifying the bulk surface term $\Theta^{\mu}$ as 
\begin{align}   \label{}
\tilde{\Theta}^{\eta}(\delta \Psi) =&~ \Theta^{\eta}(\delta \Psi) + \delta(2\sqrt{-\gamma}L_{GH}) + \delta(\sqrt{-\gamma}L_{ct}) =  \sqrt{-\gamma}\Big(T^{ij}_{B}\delta \gamma_{ij} + \Pi_{\psi}\delta \psi\Big)\,,  
\end{align}
where the second line equality  comes from Eq.~(\ref{ReAction}). Holographic renormalization condition and
$\tilde{\Theta}$-expression tells us that $\tilde{\Theta}^{\eta} \sim \CO(1)$ in the radial expansion.
Correspondingly, the modified on-shell Noether current $\tilde{J}^{\eta}$ for a diffeomorphism parameter $\zeta$ becomes
\begin{equation} \label{ModRel}
\tilde{J}^{\eta} = \p_{i}\tilde{K}^{\eta i}(\zeta) = \zeta^{\eta}\sqrt{-\gamma}\CL^{on}_{r} - \tilde{\Theta}^{\eta}(\Lie_{\zeta} \Psi)\,,
\end{equation}
where we have used the on-shell condition on the bulk background fields. 
Just as in the case of $\delta \xi^{\mu}=0$~\cite{Hyun:2014sha,Papadimitriou:2005ii},  the asymptotic behavior of general diffeomorphism parameter $\zeta$  is given by 
$\zeta^{\eta} \sim \CO(e^{-d\eta})$ and $\zeta^{i} \sim \CO(1)$,
in order to preserve the asymptotic gauge choice and the renormalized action.
This asymptotic behavior in the diffeomorphism parameter $\zeta$ allows us to discard the first term in the right hand side of Eq.~(\ref{ModRel}) when we approach the boundary. In the following we keep only the relevant boundary values of parameters such that  a bulk Killing vector $\xi^{i}$ is replaced by its boundary value $\xi^{i}_{B}$.
For the diffeomorphism variation $\Lie_{\zeta}\Psi$, the modified surface term $\tilde{\Theta}^{\eta}$ becomes
\begin{equation} \label{}
\tilde{\Theta}^{\eta}(\Lie_{\zeta}\Psi) = \sqrt{-\gamma}\Big(2T^{ij}_{B}\nabla_{i}\zeta_{j} + \Pi_{\psi}\Lie_{\zeta}\psi\Big) = \p_{i}\Big(2\sqrt{-\gamma}\,{\bf T}^{ij}_{B} \,\zeta_j\Big)\,, \nn
\end{equation}
where we have used the identity given in Eq.~(\ref{Bid}).  By using this result, one can see  that the Noether potential $\tilde{K}^{\eta i}$, up to the irrelevant total derivative term, is given by
$
\tilde{K}^{\eta i} = -2\sqrt{-\gamma}\,{\bf T}^{ij}_{B} \,\zeta_j$.

As a result, the on-shell relation between the ADT and Noether potentials for a Killing vector $\xi_{B}$ is given by
\begin{align}   \label{equivalence}
\sqrt{-g}Q^{\eta i}_{ADT}& |_{\eta\rightarrow \infty}  = \sqrt{-\gamma}\CJ^{i}_{B}\,. 
\end{align}
This shows us the scheme independence of the holographic charges since their currents are identified with covariant bulk ADT potentials which are regardless of the counter terms. We would like to emphasize that the above potential-current relation holds up to the total derivative terms which are irrelevant in the charge computation. Moreover this equality guarantees the Smarr relation since the relation was shown to hold in bulk formalisms~\cite{Barnich:2004uw,Hyun:2014kfa}.

Since we have presented formal arguments, it would be illuminating to show the frame and scheme independence of mass and angular momentum of five-dimensional AdS Kerr black holes as an explicit example, which is done  in the following section.

\section{Five-dimensional example}

As a specific example, let us focus on the pure Einstein gravity on five dimensions. In the following we will set the radius of the asymptotic AdS space as unity, $L=1$. 
AdS Kerr black hole solutions in  Boyer-Lindquist coordinates~\cite{Hawking:1998kw} are given by 
\begin{align}   \label{}
ds^{2}=&~ -\frac{\Delta_{r}}{\rho^{2}}\Big(dt - a \Delta_{\phi}d\phi -  b\Delta_{\psi}d\psi\Big)^{2}  + \frac{\rho^{2}}{\Delta_{r}}dr^{2} + \frac{\rho^{2}}{\Delta_{\theta}}d\theta^{2} \nn \\
&~+ \frac{\Delta_{\theta}\sin^{2}\theta}{\rho^{2}}\Big(adt - \frac{r^{2}+a^{2}}{1-a^{2}}d\phi\Big)^{2} + \frac{\Delta_{\theta}\cos^{2}\theta}{\rho^{2}}\Big(bdt - \frac{r^{2}+b^{2}}{1-b^{2}}d\psi\Big)^{2}   \\
&~+ \frac{1+1/r^{2}}{\rho^{2}}\Big( abdt -  b(r^{2}+a^{2})\Delta_{\phi} d\phi -a(r^{2}+b^{2}) \Delta_{\psi}d\psi \Big)^{2}\,,   \nn 
\end{align}
where  $ \rho^{2} \equiv r^{2}+a^{2}\cos^{2}\theta + b^{2}\sin^{2}\theta$, 
\begin{align}   \label{BLcoord}
 \Delta_{r}&\equiv (r^{2}+a^{2})(r^{2}+b^{2})\Big(1+ \frac{1}{r^{2}}\Big) - 2m\,, \nn\\
\Delta_{\theta} & \small \equiv 1-a^{2}\cos^{2}\theta -b^{2}\sin^{2}\theta\,, \quad  \Delta_{\phi} \equiv \frac{\sin^{2}\theta}{1-a^{2}}\,, \quad \Delta_{\psi}\equiv \frac{\cos^{2}\theta}{1-b^{2}}  \,.\nn
\end{align}
In order to use the holographic method, it is useful to take the radial expansion of the metric in Fefferman-Graham coordinates as
\begin{equation} \label{}
ds^{2}=   d\eta^{2} + \gamma_{ij}dx^{i}dx^{j}\,, \quad \gamma_{ij} = \sum_{n=0} e^{-2(n-1)\eta}\gamma^{(n)}_{ij}\,, ~~
\end{equation}
where the non-vanishing components of background metric $\gamma^{(0)}$ are given by
\begin{align} \label{}
\gamma^{(0)}_{tt} = -1\,,\quad  \gamma^{(0)}_{t\phi}=a \Delta_{\phi}\,,\quad    \gamma^{(0)}_{t\psi}=  b \Delta_{\psi}\,,\quad  \gamma^{(0)}_{\theta\theta} =\frac{1}{\Delta_{\theta}}\,,\quad \gamma^{(0)}_{\phi\phi}= \Delta_{\phi}\,,\quad  \gamma^{(0)}_{\psi\psi} = \Delta_{\psi}\,. \nn
\end{align}
In the computation of conserved charges, it turns out that the expansion up to the second order is sufficient. 
The non-vanishing components of the first order $\gamma^{(1)}$ are given  by
\begin{align}   \label{}
\gamma^{(1)}_{tt} &= - \frac{1}{2}(a^{2}+b^{2}+\Delta_{\theta})\,, \quad  \gamma^{(1)}_{t\phi}=\frac{a\Delta_{\phi}}{2}\big(a^{2}-b^{2}-\Delta_{\theta}\big)\,, \quad  \gamma^{(1)}_{t\psi} =  \frac{b\Delta_{\psi}}{2}\big(b^{2}-a^{2}-\Delta_{\theta}\big)\,, \nn \\
  \gamma^{(1)}_{\theta\theta} &=\frac{(2-a^{2}-b^{2}-3\Delta_{\theta})}{2\Delta_{\theta}}\,, \quad \gamma^{(1)}_{\phi\phi} = \frac{\Delta_{\phi}}{2}\big(a^{2}-b^{2}-\Delta_{\theta}\big)\,,  \quad   \gamma^{(1)}_{\psi\psi} = \frac{\Delta_{\psi}}{2}\big(b^{2}-a^{2}-\Delta_{\theta}\big)\,,  \nn 
\end{align}
and those of the second order $\gamma^{(2)}$ are 
\begin{align}
\gamma^{(2)}_{tt}&= 3 m  - \frac{1}{8} (a^2 - b^2)^2 -  \frac{1}{4}(2 - a^2 - b^2) \Delta_{\theta} + \frac{3}{8}\Delta_{\theta}^2\,, \nn\\
\gamma^{(2)}_{t\phi} &= a \Delta_{\phi}\Big[-3 m   + \frac{1}{8}  (a^2 - b^2)^2 - \frac{1}{4} (a^2 - b^2)\Delta_{\theta} + \frac{1}{8}\Delta_{\theta}^2\Big]\,, \nn\\ 
\gamma^{(2)}_{t\psi}&=b \Delta_{\psi}\Big[-3 m   + \frac{1}{8}  (a^2 - b^2)^2 - \frac{1}{4} (b^2 - a^2)\Delta_{\theta} + \frac{1}{8}\Delta_{\theta}^2\Big]\,, \nn\\
\gamma^{(2)}_{\theta\theta}&= \small \frac{1}\Delta_{\theta}\Big[m   + \frac{(2 - a^2 - b^2)^2}{8}   - 
 \frac{3\Delta_{\theta} }{4} (2 - a^2 - b^2)+ \frac{9\Delta_{\theta}^2}{8}\Big]\,, \nn \\
\gamma^{(2)}_{\phi\phi}&= \Delta_{\phi}\Big[ m\big(1+ 4a^{2}\Delta_{\phi}\big)+\frac{(a^2 - b^2)^2}{8}  -\frac{(a^2 - b^2) \Delta_{\theta} }{4} +  \frac{\Delta^{2}_{\theta}}{8} \Big] \,, \nn\\
\gamma^{(2)}_{\psi\psi}&= \Delta_{\psi}\Big[m\big(1+ 4b^{2}\Delta_{\psi}\big)+\frac{(a^2 - b^2)^2}{8}  -\frac{(b^2 - a^2) \Delta_{\theta}}{4}  +  \frac{\Delta^{2}_{\theta}}{8}   \Big]  \,,  \nn  \\
 \gamma^{(2)}_{\phi\psi}&= 4abm\Delta_{\phi}\Delta_{\psi}\,. \nn 
\end{align}

Now, it is straightforward to obtain the expression of $\sqrt{-\gamma} \CJ^{i}_{B} (\xi_B)$ by using Eq.~(\ref{Bcur1}). Since the first term in Eq.~(\ref{Bcur1}) was already given  in~\cite{Papadimitriou:2005ii}, let us focus on the second and third terms. One may recall that the time-like Killing vector in this metric is given by
 $\xi_{T}^{i}\p_{i} = \p_{t} -a \p_{\phi} - b\p_{\psi}$.
After some computations\cite{xAct} with $0\le \theta <\frac{\pi}{2}$, $0 \le \phi, \psi < 2\pi$, it turns out that 
\begin{align}   \label{}
 \int  d^{3} x_{i} \sqrt{-\gamma}  \Big[
 {\bf T}^{i}_{\!B\, j}\delta \xi^{j}_{ T} +& \frac{1}{2}\, \xi^{i}_{T}\big(T^{kl}_{B}\delta \gamma_{kl} + \Pi_{\psi}\delta \psi\big) \Big] \nn \\
&= -\frac{\pi^{2}(a^{2}-b^{2})(2-a^{2}-b^{2})}{6(1-a^{2})(1-b^{2})}\bigg[ \frac{a\delta a}{1-a^{2}} - \frac{b\delta b}{1-b^{2}}\bigg] \,,
\end{align}
%
which results in the linearized mass expression of AdS Kerr black holes from the boundary current as 
\begin{align}   \label{}
\delta M &= \delta Q_{B}(\xi_{T}) \nn \\
&=  \frac{\pi}{2G} \bigg[\frac{ ma\delta a (5-a^{2}-3b^{2}-a^{2}b^{2})}{(1-a^{2})^{3}(1-b^{2})^{2}} + \frac{mb\delta b (5-b^{2}-3a^{2}-a^{2}b^{2})}{(1-a^{2})^{2}(1-b^{2})^{3}}+ \frac{\delta m (3-a^{2}-b^{2} -a^{2}b^{2})}{2(1-a^{2})^{2}(1-b^{2})^{2}} \bigg]\,.  \nn
\end{align}
One can check that the difference between  our  mass expression of $\delta M$ and the conventional one in~\cite{Papadimitriou:2005ii} resides only in absence of the rotational parameter dependence of Casimir energy part. The finite mass expression is given by
\begin{align}   \label{}
M = \frac{3 \pi}{32 G} + \frac{ \pi m (3-a^{2}-b^{2} -a^{2}b^{2})}{4G(1-a^{2})^{2}(1-b^{2})^{2}} \,, 
\end{align}
where we have added the constant Casimir energy part as an integration constant. For rotational Killing vectors  $\xi_{R1}^{\mu}\p_{\mu} = -\p_{\phi}$ and $\xi_{R2}^{\mu}\p_{\mu} = -\p_{\psi}$, one can see that the additional terms, {\it i.e.} second and third ones in Eq.~(\ref{Bcur1}), vanish and so the angular momentum expressions are identical with those given in~\cite{Papadimitriou:2005ii}, which is also the case in the computation of Wald's entropy of black holes.

Now, let us check the frame independence for our expression by considering  different coordinates. 
In asymptotically canonical AdS coordinates, the metric of AdS Kerr black holes can be taken in the form of~\cite{Gibbons:2005jd} 
\begin{align}
	ds^{2} =& -(1+y^{2})dt^{2} + \frac{dy^{2}}{1+y^{2}-\frac{2m}{\Delta_{\hat{\theta}}^{2}y^{2}}} + y^{2}d\hat{\Omega}_{3}^{2} \\
	&+ \frac{2m}{\Delta_{\hat{\theta}}^{3}y^{2}}(dt - a\sin^{2}\hat{\theta}d\hat{\phi} - b\cos^{2}\hat{\theta}d\hat{\psi})^{2} + \cdots\,, \nn
\end{align}
where 
\begin{align}
\Delta_{\hat{\theta}} &\equiv 1 - a^{2}\sin^{2}\hat{\theta} - b^{2}\cos^{2}\hat{\theta}\,,\nn\\
	d\hat{\Omega}_{3}^{2} &\equiv d\hat{\theta}^{2} + \sin^{2}\hat{\theta}d\hat{\phi} + \cos^{2}\hat{\theta}d\hat{\psi}\,.\nn
\end{align}
By using  Fefferman-Graham coordinates,  
one can check  explicitly that mass and angular momentums in these non-rotating coordinates  are given by the same expressions as in the rotating ones. (See also~\cite{Gibbons:2005jd}.)

For comparison, let us turn to the bulk covariant expressions of ADT potentials. In Einstein gravity, the Noether potential $K^{\mu\nu}$ and the bulk surface term $\Theta^{\mu}$ can be taken respectively as
$K^{\mu\nu}(g\,;\, \zeta) = 2\nabla^{[\mu} \zeta^{\nu] }$ and $\Theta^{\mu} (g\,;\,\delta g)= 2\sqrt{-g} g^{\alpha[\mu} \nabla^{\beta]}\delta g_{\alpha\beta}$.
The ADT potential, $Q^{\mu\nu}_{ADT}(\xi_{T}\,;\,\delta a, \delta b, \delta m)$ for  AdS Kerr black holes is composed of three terms which correspond to the variations of parameters $a$, $b$ and $m$, respectively as  $Q^{\mu\nu}_{ADT}(\xi_{T}\,;\,\delta m)$, $Q^{\mu\nu}_{ADT}(\xi_{T}\,;\,\delta a)$ and $Q^{\mu\nu}_{ADT}(\xi_{T}\,;\,\delta b)$.

 For the bulk Killing vector $\xi_{T}$ taken in the same form as the boundary time-like Killing vector, 
the relevant component of the $Q^{\mu\nu}_{ADT}(\xi_{T}\,;\,\delta m)$ term
is given by
\begin{align}
&2\sqrt{-g}Q^{\eta t}_{ADT} (\xi_{T}\,;\,\delta m) =  \frac{-\delta m\, \sin2\theta}{(1-a^{2})^{2}(1-b^{2})^{2}}\Big[(a^{2}+b^{2}+a^{2}b^{2}-3) +2 (a^{2}-b^{2})\cos 2\theta\Big]\,.  \nn
\end{align}
The relevant component of the $Q^{\mu\nu}_{ADT}(\xi_{T}\,;\,\delta a)$ term is given by
\begin{align}   \label{}
 2\sqrt{-g} Q^{\eta t}_{ADT} (\xi_{T}\,;\,\delta a) & =   \frac{-a\delta a\, \sin2\theta}{(1-a^{2})(1-b^{2})} \bigg[\frac{(b^{2}-a^{2})   }{8} +  \frac{2m(-5+3b^{2} + a^{2} + a^{2}b^{2})}{(1-a^{2})^{2}(1-b^{2})}   \nn \\
& \qquad\qquad\quad   +\Big\{ \frac{1}{2}(2-a^{2}-b^{2}-4e^{2\eta}) + \frac{2m(1-3(b^{2}-a^{2}) -a^{2}b^{2})}{(1-a^{2})^{2}(1-b^{2})}\Big\} \cos2\theta  \nn \\
& \qquad\qquad\quad + \frac{3}{8}(b^{2}-a^{2})\cos4\theta\bigg]\,, \nn 
\end{align}
where one may note that the potentially divergent term proportional to $e^{2\eta}$  corresponds to the irrelevant  total derivative one.  $Q^{\mu\nu}_{ADT}(\xi_{T};\delta b)$ is given just by exchanging  $(a, \delta a)$ by $(b, \delta b)$ in  the above $Q^{\mu\nu}_{ADT}(\xi_{T};\delta a)$ expression.   One may note that the varying Killing vector  contribution in Eq.~(\ref{bulkADT}) does not vanish and is given by
\[
	K^{\eta t}(\delta\xi_{T})=\frac{8ma\cos\theta\sin^{3}\theta}{(1-a^{2})^{2}(1-b^{2})}\delta a+ \frac{8mb\cos^{3}\theta\sin\theta}{(1-a^{2})(1-b^{2})^{2}}\delta b\,. \]
%
%
%
%
%
%
Now, it is straightforward to check the matching between the linearized mass expression of AdS Kerr black holes  as
\begin{equation} \label{}
\delta M_{ADT} = \frac{1}{16\pi G}\int d\theta d\phi d\psi\, 2\sqrt{-g} Q^{\eta t}_{ADT} = \delta M\,,
\end{equation}
It is also straightforward to obtain the ADT potentials  for  rotational Killing vectors and check its equivalence with the results from the boundary currents. 
%
%
%
%
%

\section{ Conclusion}

In this paper, we have proposed  how to modify  the conventional expression of holographic conserved charges in order to give the identical results with those from bulk formalisms. Our construction of holographic charges is based on the conserved boundary current, of which form is motivated by the off-shell extension of the traditional ADT formalism for bulk charges. This boundary current is composed of two parts, one of which corresponds to the conventional expression of holographic charges and the other of which does to the additional terms compensating the frame and scheme dependence of the first term. 
We would like to emphasize that our modification of holographic charge expression does not mean the change of the conventional AdS/CFT dictionary for boundary stress tensor. Rather, our modification corresponds to another prescription, in the gravity context, of holographic charge construction from boundary stress tensor in such a way that it does not depend on the frames for the asymptotic AdS space. In the bulk side, we have extended our previous covariant construction of quasi-local conserved charges when Killing vectors are varied under a generic variation. By showing the equivalence of the modified holographic expression of conserved charges to the bulk covariant expression, we have argued the consistency of our holographic expression with the standard form of the first law of black hole thermodynamics and the Smarr relation. Through the example, it is explicitly shown  that the boundary-bulk equivalence is satisfied  up to  the irrelevant total derivative term. It is also shown that the additional terms in the boundary current vanish in the case of the angular momentum and black hole entropy computation, while these remove the frame-dependence in the mass computation.

Since our boundary and bulk constructions of conserved charges are based on a single formalism which depends only on the Euler-Lagrange expression of the given Lagrangian, our construction can be presented in the unified manner and seems very natural. Furthermore,
our bulk construction is completely consistent with the well-known formalisms. In all, various constructions are naturally connected and their relationships are revealed in a unified way.  It would be very interesting to generalize our construction to the case of more general asymptotic boundary space.

\vskip 1cm
\centerline{\large \bf Acknowledgments}
\vskip0.5cm

{SH was supported by the National Research Foundation of Korea(NRF) grant funded 
by the Korea government(MOE) with the grant number  2012046278 and the grant number NRF-2013R1A1A2011548.  S.-H.Yi was supported by the National Research Foundation of Korea(NRF) grant funded by the Korea government(MOE) (No.  2012R1A1A2004410). J. Jeong was supported by the research grant ``ARISTEIA II", 3337 ``Aspects of three-dimensional CFTs", by the Greek
General Secretariat of Research and Technology.}

\section*{Appendix: Some formulae}
\renewcommand{\theequation}{A.\arabic{equation}}
  \setcounter{equation}{0}
In order to verify the conservation of boundary currents, let us start from the double variation of fields and actions. When the diffeomorphism parameter $\zeta$ is varied under a generic variation, the variation of any quantity $F^{\mu\nu\cdots}$ containing $\zeta$  is taken as
$\delta F^{\mu\nu\cdots}(\zeta\,;\, \Psi) \equiv F^{\mu\nu\cdots}(\zeta+\delta \zeta\,;\, \Psi + \delta \Psi) - F^{\mu\nu\cdots}(\zeta\,;\, \Psi )$.
For instance, the Killing conditions for the background field $\Psi$ and the varied field $\Psi+\delta \Psi$  are given respectively by$
\Lie_{\xi}\Psi=0$ and  $\Lie_{\xi +\delta \xi}(\Psi + \delta \Psi) =0$. 
When a diffeomorphism parameter is transformed under a variation such that $\delta \zeta^{\mu} \neq 0$, one needs to modify the commutation of two generic variations as
\begin{equation} \label{}
(\delta\delta_{\zeta} - \delta_{\zeta}\delta) \Psi = \delta_{\delta \zeta}\Psi\,, \qquad (\delta \delta_{\zeta} - \delta_{\zeta}\delta)I[\Psi]= \delta_{\delta \zeta}I[\Psi]\,. \nn
\end{equation}
For a boundary Killing vector $\xi_{B}$, one can see that
\begin{align}   \label{dvar}
(\delta_{\xi_{B}}\delta -\delta \delta_{\xi_{B}}) I^{on}_{r} [\Psi_{B}] &= \frac{1}{16\pi G}\int d^{d}x~ \delta_{\xi_{B}}\Big[\sqrt{-\gamma} \Big(T^{ij}_{B}\delta \gamma_{ij} +\Pi_{\psi}\delta \psi\Big) \Big] \nn \\
&= \frac{1}{16\pi G}\int d^{d}x~ \p_{i}\Big[\xi^{i}_{B}\sqrt{-\gamma}\Big(T^{kl}_{B}\delta \gamma_{kl} +\Pi_{\psi}\delta \psi\Big) \Big]\,,
\end{align}
where we have used $\delta_{\xi_{B}}\Psi_{B}=0$ and thus $\delta_{\xi_{B}}I^{on}_{r}[\Psi_{B}]=0$ in the first equality and $\delta_{\xi_{B}}=\Lie_{\xi_{B}}$ in the second equality.  The variation with respect to $\delta \xi^{i}_{B}$ can be written as
\begin{align} \label{xivar}
\delta_{\delta{\xi_{B}}}I^{on}_{r}[\Psi_{B} ] &= \frac{1}{16\pi G}\int d^{d}x \sqrt{-\gamma}\Big(-2\,T^{B}_{ij}\nabla^{i}\delta \xi^{j}_{B} +\Pi_{\psi}\Lie_{\xi_{B}} \psi\Big)  \nn \\
&= \frac{1}{16\pi G}\int d^{d}x\, \p_{i}\Big(-2\sqrt{-\gamma}\,{\bf T}^{i}_{\!B\, j}\delta \xi^{j}_{B}\Big) \,,
\end{align}
where we have used  the identity Eq.~(\ref{Bid}) in the second equality.
By identifying Eq.~(\ref{dvar}) and Eq.~(\ref{xivar}),  one can finally see that
\begin{equation} \label{}
\nabla_{i}\left[{\bf T}^{i}_{\!B\, j}\delta \xi^{j}_{B}  + \frac{1}{2}\, \xi^{i}_{B}\Big(T^{kl}_{B}\delta \gamma_{kl} + \Pi_{\psi}\delta \psi\Big)\right]=0\,.
\end{equation}
%





\begin{thebibliography}{99} 


\bibitem{Balasubramanian:1999re} 
  V.~Balasubramanian and P.~Kraus,
  ``A Stress tensor for Anti-de Sitter gravity,''
  Commun.\ Math.\ Phys.\  {\bf 208}, 413 (1999)
  [hep-th/9902121].


\bibitem{Brown:1992br} 
  J.~D.~Brown and J.~W.~York, Jr.,
  ``Quasilocal energy and conserved charges derived from the gravitational action,''
  Phys.\ Rev.\ D {\bf 47}, 1407 (1993)
  [gr-qc/9209012].


\bibitem{Hollands:2005wt} 
  S.~Hollands, A.~Ishibashi and D.~Marolf,
  ``Comparison between various notions of conserved charges in asymptotically AdS-spacetimes,''
  Class.\ Quant.\ Grav.\  {\bf 22}, 2881 (2005)
  [hep-th/0503045].


\bibitem{Papadimitriou:2005ii} 
  I.~Papadimitriou and K.~Skenderis,
  ``Thermodynamics of asymptotically locally AdS spacetimes,''
  JHEP {\bf 0508}, 004 (2005)
  [hep-th/0505190].


\bibitem{Hollands:2005ya} 
  S.~Hollands, A.~Ishibashi and D.~Marolf,
  ``Counter-term charges generate bulk symmetries,''
  Phys.\ Rev.\ D {\bf 72}, 104025 (2005)
  [hep-th/0503105].


\bibitem{Awad:1999xx} 
  A.~M.~Awad and C.~V.~Johnson,
  ``Holographic stress tensors for Kerr - AdS black holes,''
  Phys.\ Rev.\ D {\bf 61}, 084025 (2000)
  [hep-th/9910040].


\bibitem{Gibbons:2005jd} 
  G.~W.~Gibbons, M.~J.~Perry and C.~N.~Pope,
  ``AdS/CFT Casimir energy for rotating black holes,''
  Phys.\ Rev.\ Lett.\  {\bf 95}, 231601 (2005)
  [hep-th/0507034].


\bibitem{deHaro:2000xn} 
  S.~de Haro, S.~N.~Solodukhin and K.~Skenderis,
  ``Holographic reconstruction of space-time and renormalization in the AdS / CFT correspondence,''
  Commun.\ Math.\ Phys.\  {\bf 217}, 595 (2001)
  [hep-th/0002230].


\bibitem{Bianchi:2001kw} 
  M.~Bianchi, D.~Z.~Freedman and K.~Skenderis,
  ``Holographic renormalization,''
  Nucl.\ Phys.\ B {\bf 631}, 159 (2002)
  [hep-th/0112119].

\bibitem{Wald:1993nt} 
  R.~M.~Wald,
  ``Black hole entropy is the Noether charge,''
  Phys.\ Rev.\ D {\bf 48}, 3427 (1993)
  [gr-qc/9307038].



\bibitem{Kim:2013zha} 
  W.~Kim, S.~Kulkarni and S.~H.~Yi,
  ``Quasilocal Conserved Charges in a Covariant Theory of Gravity,''
  Phys.\ Rev.\ Lett.\  {\bf 111}, no. 8, 081101 (2013)
  [arXiv:1306.2138 [hep-th]].


\bibitem{Kim:2013cor} 
  W.~Kim, S.~Kulkarni and S.~H.~Yi,
  ``Quasilocal conserved charges in the presence of a gravitational Chern-Simons term,''
  Phys.\ Rev.\ D {\bf 88}, no. 12, 124004 (2013)
  [arXiv:1310.1739 [hep-th]].


\bibitem{Hyun:2014kfa} 
  S.~Hyun, S.~A.~Park and S.~H.~Yi,
  ``Quasi-local charges and asymptotic symmetry generators,''
  JHEP {\bf 1406}, 151 (2014)
  [arXiv:1403.2196 [hep-th]].


\bibitem{Gim:2014nba} 
  Y.~Gim, W.~Kim and S.~H.~Yi,
  ``The first law of thermodynamics in Lifshitz black holes revisited,''
  JHEP {\bf 1407}, 002 (2014)
  [arXiv:1403.4704 [hep-th]].


\bibitem{Hyun:2014sha} 
  S.~Hyun, J.~Jeong, S.~A.~Park and S.~H.~Yi,
  ``Quasi-local conserved charges and holography,''
  arXiv:1406.7101 [hep-th].


\bibitem{Abbott:1981ff} 
  L.~F.~Abbott and S.~Deser,
  ``Stability of Gravity with a Cosmological Constant,''
  Nucl.\ Phys.\ B {\bf 195}, 76 (1982).


\bibitem{Abbott:1982jh} 
  L.~F.~Abbott and S.~Deser,
  ``Charge Definition in Nonabelian Gauge Theories,''
  Phys.\ Lett.\ B {\bf 116}, 259 (1982).


\bibitem{Deser:2002rt} 
  S.~Deser and B.~Tekin,
  ``Gravitational energy in quadratic curvature gravities,''
  Phys.\ Rev.\ Lett.\  {\bf 89}, 101101 (2002)
  [hep-th/0205318].


\bibitem{Deser:2002jk} 
  S.~Deser and B.~Tekin,
  ``Energy in generic higher curvature gravity theories,''
  Phys.\ Rev.\ D {\bf 67}, 084009 (2003)
  [hep-th/0212292].


\bibitem{Skenderis:2002wp} 
  K.~Skenderis,
  ``Lecture notes on holographic renormalization,''
  Class.\ Quant.\ Grav.\  {\bf 19}, 5849 (2002)
  [hep-th/0209067].


\bibitem{Lee:1990nz} 
  J.~Lee and R.~M.~Wald,
  ``Local symmetries and constraints,''
  J.\ Math.\ Phys.\  {\bf 31}, 725 (1990).



\bibitem{Iyer:1994ys} 
  V.~Iyer and R.~M.~Wald,
  ``Some properties of Noether charge and a proposal for dynamical black hole entropy,''
  Phys.\ Rev.\ D {\bf 50}, 846 (1994)
  [gr-qc/9403028].


\bibitem{Barnich:2004uw} 
  G.~Barnich and G.~Compere,
  ``Generalized Smarr relation for Kerr AdS black holes from improved surface integrals,''
  Phys.\ Rev.\ D {\bf 71}, 044016 (2005)
  [Erratum-ibid.\ D {\bf 73}, 029904 (2006)]
  [gr-qc/0412029].


\bibitem{Hawking:1998kw} 
  S.~W.~Hawking, C.~J.~Hunter and M.~Taylor,
  ``Rotation and the AdS / CFT correspondence,''
  Phys.\ Rev.\ D {\bf 59}, 064005 (1999)
  [hep-th/9811056].

\bibitem{xAct}
We have computed these using xAct packages given in $[$URL=http://www.xact.es$]$.

 
\end{thebibliography}
\end{document}